\documentclass[journal,transmag,twoside]{IEEEtran}

\usepackage{cite}
\usepackage[dvipsnames]{xcolor}
\usepackage{soul}
\usepackage{url}
\usepackage{graphicx}
\usepackage{amsmath}

\newtheorem{proposition}{Proposition}[section]

\usepackage{lipsum}
\usepackage{nicefrac}
\usepackage{textcomp}
\usepackage{gensymb}
\usepackage{multirow}
\usepackage[normalem]{ulem}

\begin{document}
\title{Security Attacks on Smart Grid Scheduling and Their Defences:\\ A Game--Theoretic Approach}

\author{\IEEEauthorblockN{M. Pilz,
F. Baghaei Naeini,
K. Grammont, 
C. Smagghe,\\
M. Davis, 
J.-C. Nebel,
L. Al-Fagih, and
E. Pfluegel}%
\thanks{M.~Pilz, F.~Baghaei Naeini, M.~Davis, J.-C.~Nebel, L.~Al-Fagih, and E.~Pfluegel are with the School of Computer Science and Mathematics, Kingston University, London, KT12EE, United Kingdom}%
\thanks{K.~Grammont and C.~Smagghe are with Normandie Universit\'{e}, ENSICAEN, UNICAEN,  CNRS,  Caen, France}%
\thanks{Corresponding author: M. Pilz (Matthias.Pilz@kingston.ac.uk)}%
\thanks{Manuscript submitted: October 2018}}

\markboth{\MakeLowercase{submitted}~October~2018}%
{Security Attacks on Smart Grid Scheduling and Their Defences}

\IEEEtitleabstractindextext{%
\begin{abstract}
The introduction of advanced communication infrastructure into the power grid raises a plethora of new opportunities to tackle climate change. This paper is concerned with the security of energy management systems which are expected to be implemented in the future smart grid. The existence of a novel class of false data injection attacks that are based on modifying forecasted demand data is demonstrated, and the impact of the attacks on a typical system's parameters is identified, using a simulated scenario. Monitoring strategies that the utility company may employ in order to detect the attacks are proposed and a game--theoretic approach is used to support the utility company's decision--making process for the allocation of their defence resources. Informed by these findings, a generic security game is devised and solved, revealing the existence of several Nash Equilibrium strategies. The practical outcomes of these results for the utility company are discussed in detail and a proposal is made, suggesting how the generic model may be applied to other scenarios.
\end{abstract}

}

\maketitle

\IEEEdisplaynontitleabstractindextext

\IEEEpeerreviewmaketitle

\section{Introduction}
\label{sec:intro}
\IEEEPARstart{D}{uring} the last decade the rise of the smart grid has shown significant potential to address not only the traditional grid problems but also support the development of power generation from renewable sources. Indeed, since electricity suppliers must meet customers' demand during peak hours, they traditionally invest in power generation capacity able to sustain those high power consumption periods. This is an expensive solution as some of those resources are only exploited sporadically. On the other hand, with the increase of greenhouse gases that impact negatively on the Earth's ecosystem, better exploitation of renewable energy sources is seen as a way to reduce their emissions~\cite{Panwar2011RoleReview}. However, their inherent intermittency and unpredictability makes their integration into the power grid particularly difficult. Therefore, management of consumption and production  plays a crucial role to facilitate power distribution as well as reduction of cost for both suppliers and consumers~\cite{Palensky2011DemandLoads}.

Traditional Demand-Side Management has designed strategies to change consumers' consumption patterns so that they better match energy generation profiles: these include peak clipping, load shifting, and flattening consumers' loads~\cite{Gellings1985TheUtilities}. Advancements in energy storage and renewable energy generation provide further opportunities to devise smarter and efficient power grids. For instance, storing energy during off-peak times eases supply during peak hours where there is high demand. Furthermore, local electricity generation reduces substantially power dissipation and transmission costs. Accordingly, the concept of `microgrids' was introduced to facilitate distribution by dividing the power grid into several smaller local grids~\cite{USDepartmentofEnergyMicrogridExchangeGroup2016}. Efficient management of these microgrids require a two-way communication system between suppliers and consumers, so that those smart grids can exploit distributed information for storage scheduling and pricing purposes~\cite{Ipakchi2009GridFuture}. 

Taking advantage of smart meters, energy storage and trading strategies, a variety of energy consumption scheduling techniques aiming at optimally distributing daily power consumption has been put forward to reduce a smart grid's peak-to-average ratio (PAR) of the aggregated load. In particular, dynamic game--theoretic frameworks have been proposed to optimise energy cost using their Nash Equilibrium~\cite{Mohsenian-Rad2010,Mohsenian-Rad2010a}. Some consider advanced battery models~\cite{Pilz2017EnergyApproach} and integrate forecasting errors~\cite{Pilz2018AErrors}. Alternatively, usage of a Stackelberg game minimising both the PAR and the system total cost has also proved promising~\cite{Soliman2014}. More generally, comprehensive reviews reveal the significant contribution that game--theoretic solutions offer in terms of reducing consumer costs and PAR values~\cite{Pilz2017,Gupta2017ChallengesReview,Sioshansi2012SmartGrid,Saad2012,Fadlullah2011}. 

Since smart grids rely on a communication network and smart meters, they may be vulnerable to cyber attacks~\cite{Mo2012Cyber-physicalInfrastructure}. As a result, appropriate defence strategies need to be put in place~\cite{Liu2009FalseGrids,Yan2012ACommunications,Wang2013CyberChallenges,Rawat2015CyberPerspectives,He2016Cyber-physicalSurvey,Tan2017SurveyApproach}. False Data Injection (FDI) is one of the most common approaches to attack cyber-physical systems~\cite{Lun2016Cyber-PhysicalStudy}. In general, FDI attacks target data integrity breaches to make profit or disturb a system. 
Since, in power grids, state estimators are the main data sources used for monitoring and controlling purposes, they are the target of data injection~\cite{Liang2017ASystems}. Such FDI attacks and possible defence strategies have been investigated in several scenarios: (i) the `ideal' undetectable attack where the attack vector is built from complete knowledge of the state estimators' parameters~\cite{Liu2009FalseGrids}; (ii) a more realistic attack relying on a probability distribution function where only incomplete information about the system's parameters is available~\cite{Rahman2012}; (iii) a stealth data injection in which an attacker has complete information about the system's topology~\cite{Huang2013}. Detection of cyber-attacks and associated defence strategies are essential for a reliable grid. For instance, a fast detection algorithm has been proposed to deal with FDI and jamming attacks in smart grids~\cite{Kurt2018Real-TimeGrid}.

Since game theory has been a very successful framework to improve cyber security~\cite{Wu2018DetectionSystems}, it has been applied in several scenarios dealing with grid security. When attackers target either a single or multiple state estimators, both Markovian and static strategies have been investigated to defend against load redistribution attacks by allocating optimal budgets to energy suppliers~\cite{Xiang2017a}. If attackers manipulate price information from the utility companies, its impact can be mittigated exploiting a Stackelberg formulation~\cite{Maharjan2013DependableApproach}. Furthermore, it has been proposed to defend against coalitional attacks by multiple attackers using an iterated game--theoretic model~\cite{Yang2016AGrid}, where a probability of attack detection is considered in each iteration: correlation  between payoffs and penalty factors demonstrated the effectiveness of the defence system. Finally, a defence system against switching attacks based on a zero-determinant iterative game between controller and attacker showed that transient stabilisation could be achieved over time~\cite{Farraj2016}. Although grid cyber security has been an active field of research, no defence scheme has yet been proposed to protect forecasting data in smart grids.

The contributions of this paper are as follows:
\begin{enumerate}
\item The design of a novel class of false data injection attacks, preserving average daily load in a smart energy scheduling system. The forecasted demand data is corrupted by a single attacker, targeting one or several households. Using extensive simulations, two families of attacks are investigated. The impact on both the PAR of the aggregated load and consumer bills as well as the resulting benefit for the attacker are analysed. 
\item The design and analysis of an augmented security game for monitoring average-preserving false data injection attacks, based on a detailed model with strategies and payoff functions informed by the simulation findings. The conditions under which a pure Nash Equilibrium exists are derived. This extends previous work by providing additional strategies and a more detailed payoff design, informed by the various cost and benefit functions of the utility company and the attacker. 
\item To give practical guidelines to the utility company on how to protect itself against such attacks. The recommendations are based on combining a range of mitigation strategies and the results of the equilibrium analysis of the game, to aid the utility company with the decision--making process of investing in the security defence. The given advice is motivated by the simulation scenario, but can also be adapted to other situations. This is demonstrated using a concrete example. 
\end{enumerate}

This paper is organised as follows. The underlying smart grid management model is introduced in Section~\ref{model_section}. Different types of attacks are developed and their impacts are analysed in Section~\ref{vulnerability_section}. A game--theoretic defence strategy for the utility company is proposed in Section~\ref{sec-game-defence}. Finally, the paper is concluded in Section~\ref{sec:conclusion}.

\section{Smart Grid Management Model}
\label{model_section}
This section focuses on the description of the game--theoretic scheduling model used in a smart grid management model. After specifying the smart grid scenario including the battery model, cost function and data specifications, the scheduling game is presented. Note that a more detailed description can be found in~\cite{Pilz2018AErrors}.   
\subsection{Scenario}
\label{sec:scenario}
The scenario of interest considers a residential neighbourhood comprised of $M$ houses where each household is equipped with a smart meter. The set of households that participates in the demand-side management (DSM) program is denoted by $\mathcal{N}\subset\mathcal{M}$, where $\mathcal{M}$ is the set of all households in the neighbourhood. The total number of participants is $N=|\mathcal{N}|$. It is assumed that all $M$ households are served by the same utility company (UC).

Each day is split into $T$ discrete intervals, where the set of all intervals is represented by $\mathcal{T}$. The DSM protocol runs as follows: The forecasted demand is sent to the UC where demand data are aggregated and sent to each DSM participant. Based on this input, the households play a dynamic non-cooperative game (cf.~Section~\ref{sec:scheduling_game}). Its outcome is a set of schedules, one for each household, that specify how they can make best use of their battery system. The households follow these schedules, even if their actual demand differs from the forecasted one. Instead of using a forecasting algorithm, random errors were added to actual demands in order to simulate a realistic average error of 8\% in individual forecasted data as reported in~\cite{Bichpuriya2016}. More details about the process used to simulate realistic forecasts can be found in Appendix~\ref{sec:forecasting}.

Households that participate in the DSM scheme are equipped with a lithium-ion battery. Using the battery model proposed in~\cite{Pilz2017EnergyApproach,Pilz2018AErrors}, which includes charging, discharging, and self-discharging characteristics of the battery, storing decisions are made. They are denoted by the variable $a$. 

The demand ${d}^t_m \geq 0$ of a household $m\in\mathcal{M}$ is defined as the amount of electricity that is needed to run all its appliances during the time interval $t\in\mathcal{T}$. Let $l_m^t$ denote the load, i.e.~the amount of energy drawn from the grid by household $m\in\mathcal{M}$ during the interval $t\in\mathcal{T}$. For the scheme participants, the load depends on the decision $a_n^t$ taken at that specific interval, it combines the demand with the amount of energy that is charged or discharged by the battery $l_n^t = d^t_n + a^{t}_n\ .$ Thus, the grid total load during interval $t$ is given by $L^t = \sum_{m\in\mathcal{M}} l_m^t\ .$

In order to incentivise a reduction of load at peak times, the UC charges the DSM participants using a dynamic pricing tariff: The cost per energy unit is based on the aggregated load of all users and is calculated separately for each interval. As in~\cite{Mohsenian-Rad2010,Yaagoubi2015b}, this is expressed by a quadratic cost function $g^t(y) = c_2\cdot y^2 + c_1\cdot y + c_0\ ,$ where $y$ is the aggregated load at time $t$ given by $L^t$ and the coefficients $c_2>0$, $c_1\geq 0$ and $c_0\geq 0$. The electricity bill $B_n$ (cf.~\cite{Pilz2017EnergyApproach,Soliman2014,Mohsenian-Rad2010}) of each participant is given by:
\begin{equation}
	B_n = -\Omega_n \sum_{t\in\mathcal{T}} g^t\ \ \forall n\in\mathcal{N}\ ,
	\label{eqn:bill}
\end{equation}
where $\Omega_n = \frac{\sum_{t}l_n^t}{\sum_{t}\sum_{k}l_k^t}\ .$

\subsection{The Scheduling Game}
\label{sec:scheduling_game}
Formally, the game used to schedule battery usage is a discrete time dynamic game, in which players, i.e.~households, have to decide how to use their battery during each individual interval of the upcoming day. In this game, each household has the objective to minimise their own costs as defined by the electricity bill \eqref{eqn:bill}. As the electricity bill itself depends on the aggregated load, the selfish behaviour of each individual becomes equivalent to minimising the peak-to-average ratio of the aggregated load (cf.~\cite{Soliman2014}). It is defined as 
\begin{equation}
	\text{PAR} = T\cdot\frac{\max_{t\in\mathcal{T}} L^t}{\sum_{t\in\mathcal{T}} L^t}\ .
    \label{eqn:PAR}
\end{equation}
The theoretically optimal result is a perfectly flat curve with a PAR value equal to 1.0. 
Since the mathematical details of the game mechanics lie outside the scope of this manuscript, the interested reader should refer to~\cite{Pilz2018AErrors} for a thorough description. In the following, the scheduling game is treated as a black--box that takes forecasted demand data as an input and then outputs schedules (one for each household) of \textit{optimal} battery usage for the upcoming day as defined by its Nash Equilibrium. An optimal Nash Equilibrium (NE) strategy has a local maximum property: any single player deviating from the NE strategy will suffer a reduced payoff. It is important to note that only unilateral strategy changes are considered in this concept. Hence applying game theory for real--life scenarios is only a valid and useful tool if all participants agree to adopt it as a contract for strategic decision--making, in the modelled scenario. 

\begin{figure}
\centering
\includegraphics[width=0.42\textwidth]{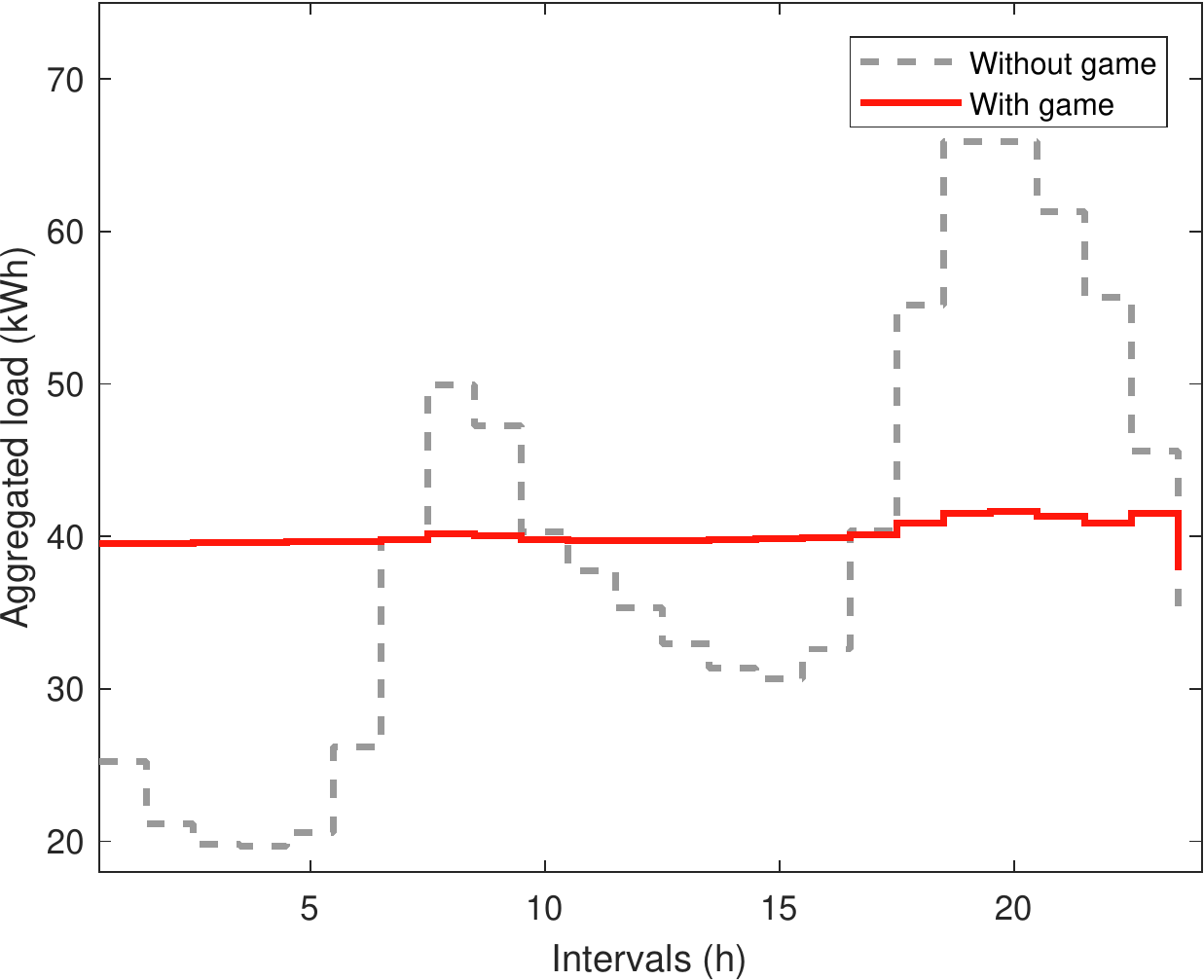}
\caption{\textit{Aggregated load comparison.} 
The aggregated load of $M=25$ households for a single day is shown for two scenarios: Without the game, and after playing the game. Every household is equipped with a battery.
As players try to lower their electricity bill~\eqref{eqn:bill} (by means of their battery usage), they directly affect the load profile. In this example, the PAR value of the aggregated load~\eqref{eqn:PAR} decreases from 1.69 to 1.04.}
\label{fig:load_game}
\end{figure}

Fig.~\ref{fig:load_game} shows an example of the scheduling impact of the game on the aggregated load for one day. Whereas the load profile without playing the game shows the usual peaks in the morning and evening, it is possible to obtain a relatively flat profile by means of the scheduling game. The first row of Fig.~\ref{fig:attackExplanation} (in Appendix~\ref{sec:supplementary}) illustrates actual battery usage of each household using a battery. As the dashed curve in the last row of Fig.~\ref{fig:attackExplanation} shows, the higher the participation to the game, the flatter the aggregated load.

\subsection{Experimental Setup}
\label{sec:setup}
Throughout the paper, all simulations are performed for a neighbourhood of $M=25$ households over a period of 365 consecutive days to allow for statistical analysis of the outcomes. Each day is split into $T=24$ intervals. The respective demand data are taken from~\cite{Openei2013a}. Every participant of the DSM scheme is equipped with the same type of battery, i.e.~the Tesla Powerwall 2 (cf.~\cite{Tesla2017}). Battery characteristics such as efficiency, capacity, charging and discharging rates, and degeneration behaviour are read off its data sheet. This setup is deliberately chosen to be similar to the one investigated in~\cite{Pilz2018AErrors,Pilz2018a} to allow for comparison of the outcomes.

\section{False Data Injection on Forecasts}
\label{vulnerability_section}
As motivated in Section~\ref{sec:intro}, the security of a smart energy system is of extreme importance and there is a lack of research on possible attacks on forecasted data. This section describes different types of potential attacks that may take advantage of the game--theoretic smart grid management model presented previously. Furthermore, outcomes of those attacks are analysed from the point of view of the attacker, the UC and the other players. Various defence strategies to detect those attacks are proposed and analysed. Finally, attack mitigation is discussed.

\subsection{Description of Attacks}
\begin{figure}
\centering
\includegraphics[width=0.42\textwidth]{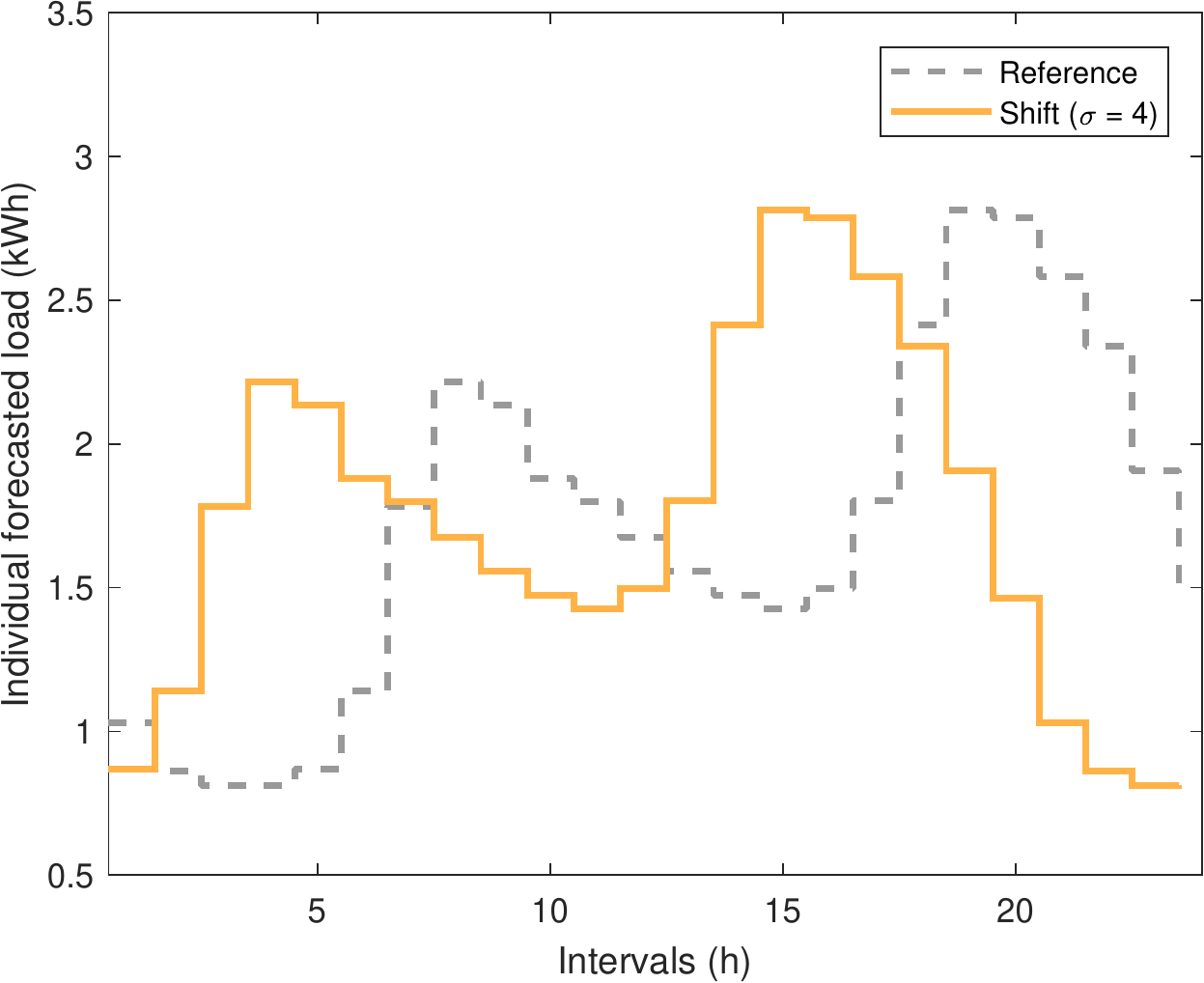}
\caption{\textit{Example of a shift attack.} The reference curve shows the forecasted load of an individual household for the upcoming day. When the attacker applies the shift attack, they perform a circular shift of the interval data. The result of a shift with $\sigma=4$ is shown as an example.}
\label{fig:shift_example}
\end{figure}

\begin{figure}
\centering
\includegraphics[width=0.42\textwidth]{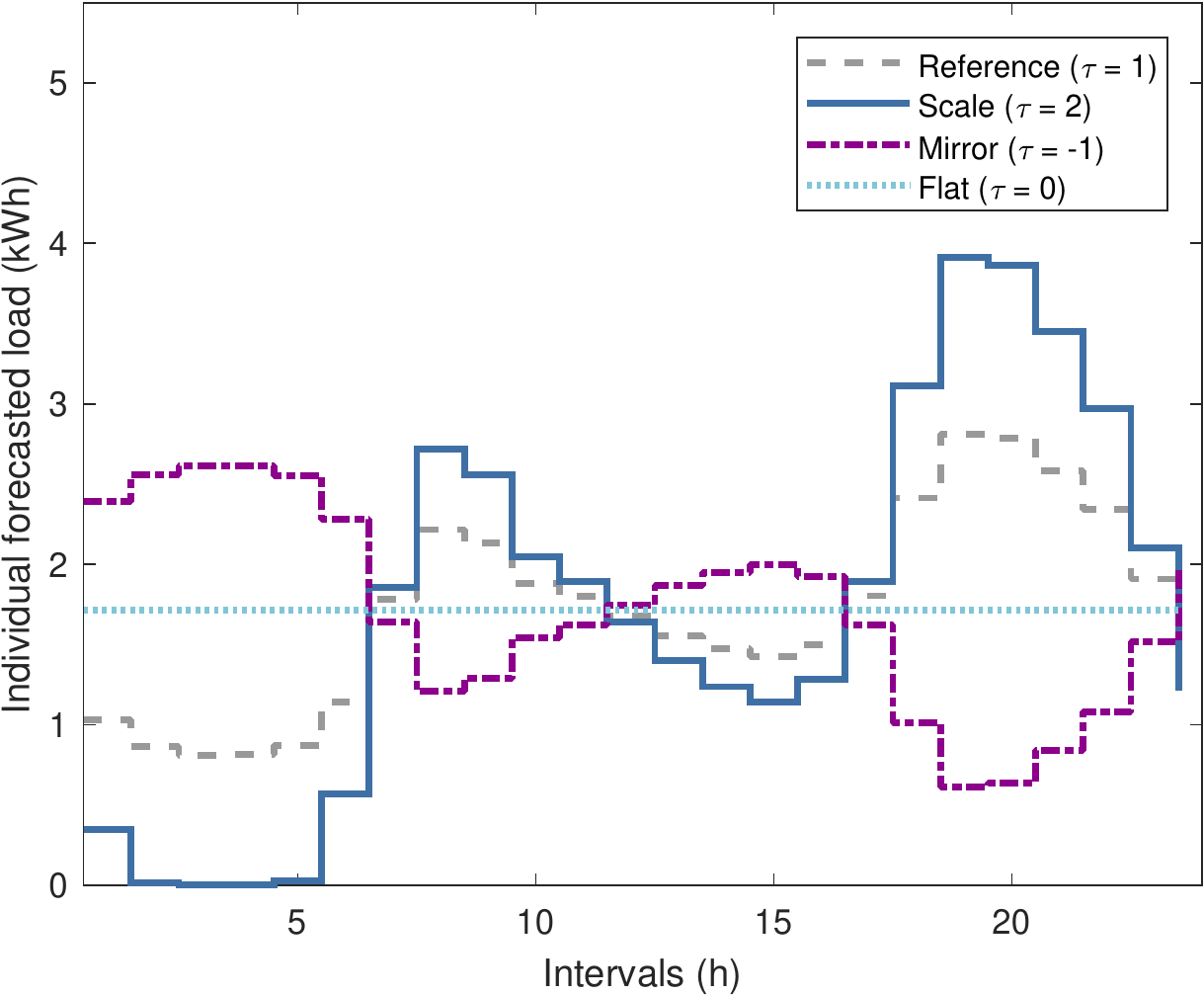}
\caption{\textit{Example of scale attacks.} The reference curve shows the forecasted load of an individual household for the upcoming day. It is identical to the reference shown in Fig.~\ref{fig:shift_example}. When the attacker applies the scale attack, they scale the interval data with respect to the daily average of the forecasted load. Scaling with a factor $\tau=2$ leads to more severe troughs and peaks, while using $\tau=-1$ results in a mirrored forecast. $\tau=0$ gives a perfectly flat load profile.}
\label{fig:scale_example}
\end{figure}

All attack scenarios investigated in this section rely on the following assumptions. First, the attacker (who is one of the players) exploits the vulnerability of the smart grid communication network: They have the ability not only to intercept forecasting data from all the other players, but also to replace them. Second, after the game has been played based on the tampered data, the attacker adapts their storage schedule and takes advantage of the erroneous schedules that the other players follow. Finally, in order to limit the risk of having their attack detected, the attacker makes sure that the average daily aggregated load is not affected by their actions. Although there are many strategies which can be applied to change forecasts while maintaining a constant aggregated value of the load, this study investigates two simple families of attacks: Forecast shifting and scaling. 
\vspace{1em}
\paragraph{Shift Attack}
The shift attack replaces a given forecast with the original forecast after having undergone a circular shift of $\sigma$ time intervals, where $\sigma$ is an integer. Since experimental results have shown that a shift attack of $4$ hours, see Fig.~\ref{fig:shift_example}, produces the most dramatic impact for the dataset of interest (cf. Section.~\ref{sec:setup}), that value is used for the rest of the study.
\vspace{1em}
\paragraph{Scale Attack}
The scale attack substitutes a given forecast with a scaled version centred around its average value for the day. To ensure that the day average is not affected, the scaling parameter $\tau$ should be chosen such that no load becomes negative after scaling. Note that for the dataset of interest (cf.~Section.~\ref{sec:setup}), a value of $\tau = 2$ remains acceptable: Although a couple of values do become negative, they are set to 0; the day average is slightly increased, but it remains within a realistic forecast uncertainty (cf.~Appendix~\ref{sec:forecasting}). Fig.~\ref{fig:scale_example} illustrates the effect of various scale attacks, i.e.~$\tau = -1$, $\tau = 0$ and $\tau = 2$. While $\tau = 1$ returns the initial forecast, $\tau = 0$ and $\tau = -1$ produces a flat, and mirrored forecast, respectively. In the rest of the paper, these two different attacks are called flat and mirror attack.

The outcome of an attack does not only depend on the type of attack and its associated parameter, but also on the number of forecasts which are replaced among all the players of a game: the higher the percentage $\rho$ of attacked households, the more room for maneuver the attacker has to profit from their attack.

\subsection{Attack Outcomes}
\subsubsection{Outcome for the Attacker}
\label{sec:outcomeForAttacker}

\begin{figure}
\centering
\includegraphics[width=0.42\textwidth]{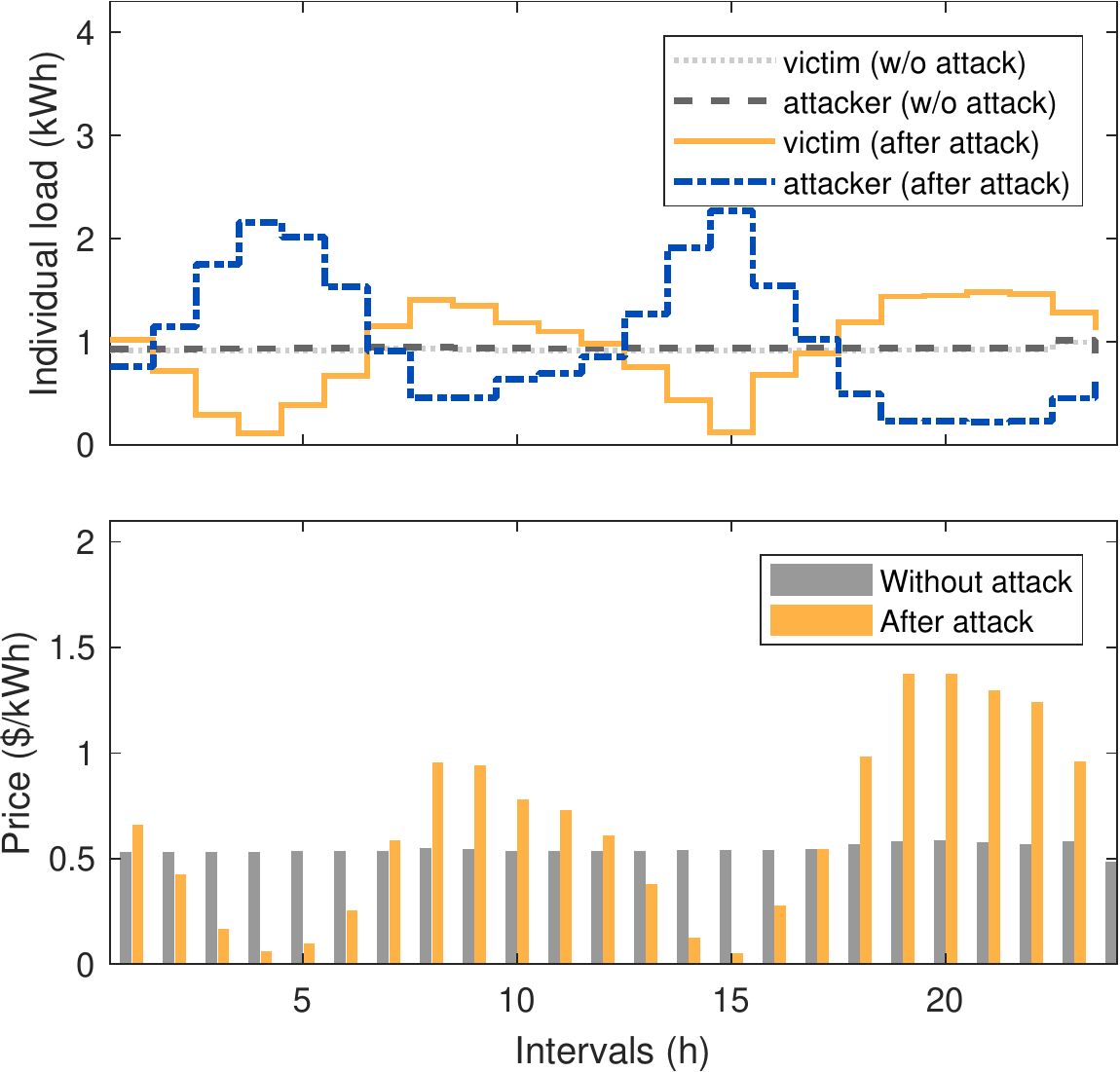}
\caption{\textit{Individual loads and unit price after scheduling.} The top graph displays load profiles for one day of a randomly picked victim and the attacker for two different scenarios: with and without attack. While both references show an almost flat profile (cf.~Fig.~\ref{fig:load_game}), the load curves after the attack differ considerably. This is a direct result of the attacker taking advantage of the falsely injected data. The bottom graph displays the change of price per unit during those two scenarios. As expected, the attacker's load has a high inverse correlation with the unit price ($\approx -0.96$).}
\label{fig:loads_attack}
\end{figure}

Fig.~\ref{fig:loads_attack} illustrates the resulting load curves of attacker and victim in the case of a shift attack ($\sigma = 4$). The attacker benefits by having a high load during the periods when the victims have a low one and vice versa, so that the attacker's higher consumption takes place when the aggregated load, and thus unit price, is low. This is exactly what the attacker tried to achieve by manipulating the forecasting data and thus the input to the scheduling game.
More details about the cost function model can be found in Section~\ref{sec:scenario} and~\cite{Rahbar2015,Pilz2017}. In this attack example, there is a high inverse correlation, i.e.~$\approx -0.96$, between the attacker's load and the unit price.
    
\begin{figure*}
\centering
\includegraphics[width = 1.0\textwidth]{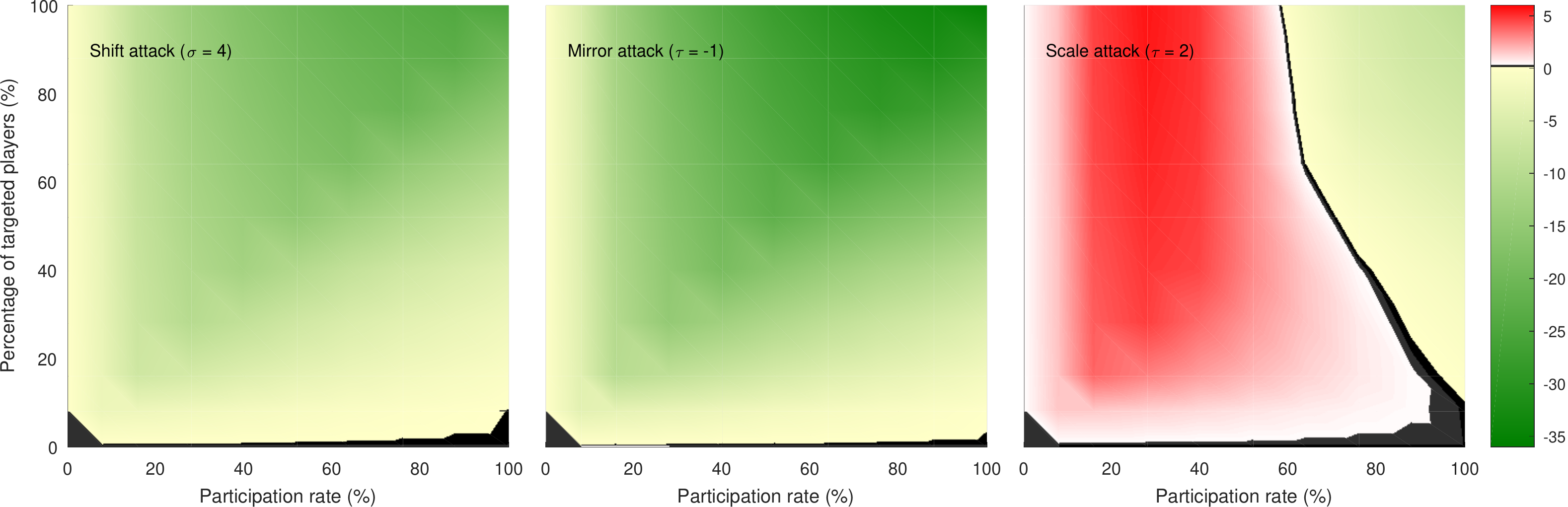}
\caption{\textit{Financial benefit for the attacker.} The median change (compared to the non--attack scenario) over 365 days of the energy bill for the attacker is shown in per cent. The outcomes for three different attacks, i.e.~shift attack with $\sigma=4$, mirror attack, and scale attack with $\tau=2$, are presented. The simulations were performed for $M=25$ using various participation rates and percentages of targeted players. While the first two attacks display similar benefits, the third one indicates that for specific scenarios the attacker also has an increased electricity bill.}
\label{fig:benefit_attacker}
\end{figure*}

An attacker's financial benefit depends not only on the type of attack, but also the number of households using a battery, i.e. the participation rate $\nicefrac{N}{M}$, as well as the proportion of targeted households $\rho$ whose forecasts have been changed. In order to investigate this, attack simulations were conducted on a smart grid comprising $M=25$ households for a duration of one year. Compared to the non-attack scenario, Fig.~\ref{fig:benefit_attacker} displays the percentage change on the attacker's bill (yearly median of the daily changes) according to those parameters in the cases of shift ($\sigma = 4$), mirror ($\tau = -1$) and scale ($\tau = 2$) attacks. Simulations have revealed that a flat attack ($\tau = 0$) results in benefits similar to those of the shift attack ($\sigma = 4$) and is thus not shown.

Fig.~\ref{fig:benefit_attacker} reveals that for shift ($\sigma = 4$) and mirror ($\tau = -1$) attacks the attacker is never penalised by their action and their gains increase with both participation rate and percentage of targeted players. Bill reductions for the attacker reach up to $25.5\%$ and $35.7\%$, respectively. However, in the case of the scale attack ($\tau = 2$), the graph displays a different picture: Up to a relatively high participation rate ($\nicefrac{N}{M}>55\%$), the attacker is financially penalised by their attack. Indeed, while the other attacks lead players to charge their battery at a wrong time, this scale attack tends to make players charge their battery more than they need at a time when the attacker would also need to charge their battery. As Fig.~\ref{fig:attackExplanation}  (cf.~Appendix~\ref{sec:supplementary}) reveals, when the participation rate is high, the aggregated load profile is inverted due to a large number of players charging their battery excessively at a time that was initially of low load and discharging their battery when a peak was expected. As a consequence, the aggregated load profile is now almost ideal for the attacker who can benefit from low prices at their time of high needs. Thus, they hardly need to use their battery and can gain up to 9.5\% of bill reduction. 
\vspace{1em}
\subsubsection{Outcome for the Utility Company and the Other Players}
\begin{figure*}
\centering
\includegraphics[width = 1.0\textwidth]{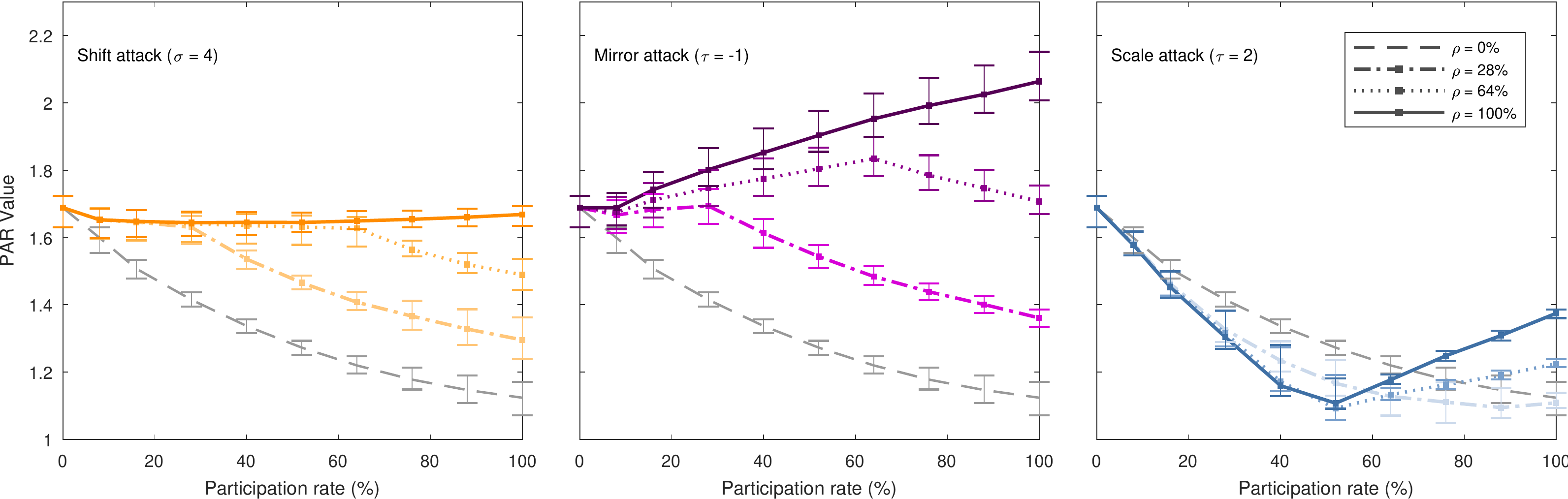}
\caption{\textit{Peak-to-average ratio (PAR) of the aggregated load for different attack scenarios.} The median PAR value for a 365 day simulation is plotted together with the range between the first and third quartile over the participation rate. The outcomes for three different attacks, i.e.~shift attack with $\sigma=4$, mirror attack, and scale attack with $\tau=2$, are presented. For each attack, the individual graphs differ in their number of attacked players. This also includes the reference outcome of the scheduling game in which no player is attacked.}
\label{fig:par_value_attack}
\end{figure*}

As mentioned in Section~\ref{model_section}, for the utility company, the efficiency of a microgrid is assessed by its PAR value. Since attacks change the aggregated load, it is directly affected. The effect of the previously introduced attacks on PAR values is presented in Fig.~\ref{fig:par_value_attack}. The different attack types are associated to a different graph, which presents several curves, each for a different percentage $\rho$ of targeted players, showing the relationship between participation rates $\nicefrac{N}{M}$ and PAR values. 
For the shift ($\sigma = 4$) and mirror ($\tau = -1$) attacks, an increase of $\rho$ leads to a worsening of PAR values. Moreover, as in the non--attack scenario, PAR values tend to improve with an increase of participation rate. Note for the case of the mirror attack: If a high percentage of players are targeted, an increase of the participation rate contributes to the degradation of PAR values. 

As analysed in the previous section, the outcomes of the scale attack ($\tau = 2$) are different from the others when the participation rate is below $\nicefrac{N}{M}=55\%$. In fact, Fig.~\ref{fig:par_value_attack} shows an improvement of the PAR values compared to the non--attack scenario when the percentage $\rho$ of targeted players increases. Fig.~\ref{fig:attackExplanation} clearly shows that at low participation rates the aggregated load is flatter than without an attack. The explanation is that this positive scaling incentivises participating households to work harder to flatten the load curve: As seen in Fig.~\ref{fig:attackExplanation}, charging takes place at the same time but with a higher intensity. As a consequence, a 52\% participation rate is sufficient to achieve a PAR that is similar to the one resulting from a 100\% participation rate without any attack, i.e.~PAR = 1.11 and PAR = 1.07, respectively. Participants work twice harder, which has the same effect as if everybody was working as they should. This extra work leads to higher bills for those households. An improved PAR value may suggest that the UC benefits from such attacks. In practice, this is not the case because in those scenarios the electricity bills of all players, including the attacker, increase substantially (data not shown), which will eventually lead to a loss of reputation and customers for the UC.

All attacks leading to the reduction of a single player's (the attacker) bill result in an increase of all the other players' bills by usually a comparable amount, see Table~\ref{tab:attacks1} and~\ref{tab:attacks2}. As a consequence, the aggregated bill for the whole neighbourhood is significantly increased.
For example, a mirror ($\tau = -1$) attack targeting all players ($\rho = 100\%$) rewards the attacker with a 35.7\% bill cut, while the other players must endure a 54.0\% rise on average. Similarly, the attacker benefits from a scale attack ($\tau = -2$, $\rho = 28\%$) with a bill reduction of 1\%, penalising the other households by a 2.3\% increase.

\subsection{Attack Detection Strategies}

All investigated attacks affect the utility company negatively: When the participation rate is high, PAR values are systematically degraded compared to the non--attack scenario; otherwise, either PAR values worsen, or their improvement is at the cost of higher electricity bills for the average household. This is detrimental to the UC's credibility and competitiveness. Consequently, the UC needs to design defence strategies to prevent attacks that affect the storage scheduling process. In this study, the focus is on the detection of false data injection by monitoring the forecasting data that are transmitted every day on the smart grid communication system. 
\vspace{1em}
\subsubsection{Attack Detection Through System Monitoring}
\label{sys_monitor}
Forecast monitoring is considered at three different levels:
\begin{itemize}
\item Aggregated consumption forecast average, i.e.~average amount monitoring
\item Aggregated consumption forecast profile, i.e.~deep aggregated monitoring
\item Household consumption forecast profiles, i.e.~deep individual monitoring
\end{itemize}
In each case, the UC would compare the received forecast data with its own estimate. While monitoring the aggregated consumption forecast total only requires the UC to forecast the daily total electricity consumption of the smart grid community as a whole, deep monitoring relies on producing hourly consumption estimates for either the entire community or each individual household. The more precise the monitoring, the more resources are needed to implement it.

Since an individual average hourly forecast error for a 24-hour period is expected to be lower than 8\%~\cite{Bichpuriya2016}, the expectation is that the difference between two forecasts, i.e. the forecast provided by the received forecast data and the forecast estimated by the UC, to be lower than twice the 8\% error of a single forecast. As a consequence, it is reasonable to assume that the UC could use a threshold of 20\% to identify an attack when using deep individual monitoring. In the case of deep aggregated monitoring, the combination of forecasts tends to lead to error reduction. As consequence, here a discrepancy of at least 10\% is used to detect an attack. Finally, since in the proposed attack scenarios, the attacker always makes sure that their attack does not change the average daily aggregated forecast, a UC relying only on average amount monitoring would not be able to detect any attack. 
Eventually, the detection of a given attack depends not only on the chosen monitoring strategy, but also the type of attack, the participation rate $\nicefrac{N}{M}$ and the percentage $\rho$ of targeted players. 
\vspace{1em}
\subsubsection{Attack Impact Analysis}
\label{sec:attackImpactAnalysis}
Based on the three proposed monitoring strategies, the consequences of undetected attacks are studied. These are evaluated by estimating an attack's impact in terms of average bill change for the attacker and the other players, bill revenue change for the UC and PAR values. Assuming a participation rate of $\nicefrac{N}{M}=100\%$, this set of experiments considers, for each attack type of interest, i.e.~shift ($\sigma = 4$), flat ($\tau = 0$), mirror ($\tau = -1$) and scale ($\tau = 2$ and $\tau = 1.29$), the most severe attack, in terms of the highest percentage $\rho$ of targeted players, that has remained undetected according to the monitoring strategy.

As Tables~\ref{tab:attacks1} and~\ref{tab:attacks2} show, all of those attacks prove beneficial to the attacker in terms of reducing their bill, while other players suffer a bill increase. Regarding the UC, it benefits financially from the general bill rise, but sees its PAR value degraded. Note that the impact of a scale ($\tau = 1.29$) attack is evaluated because it is the most powerful scale attack which can target all players ($\rho = 100\%$) without being detected by any of the proposed monitoring strategies. 
\begin{table*}
\centering
\caption{Impact of undetected attacks despite average amount monitoring. Results show median values over 365-day simulations together with their respective interquartile range. The  participation rate is $\nicefrac{N}{M}=100\%$.}
\label{tab:attacks1}
\begin{tabular}{c|c|r|r|r|r}
\hline
\multirow{2}{*}{\textbf{Attack type}} & \multirow{2}{*}{\boldmath$\rho\ (\%)$} & \multicolumn{1}{c|}{\textbf{Attacker}} & \multicolumn{1}{c|}{\textbf{Other players}} & \multicolumn{2}{c}{\textbf{Utility company}}  \\ 
& & \textbf{Bill change (\%)} & \textbf{Bill change (\%)} & \textbf{Revenues change (\%)} & \textbf{PAR value} \\ \hline\hline
Shift ($\sigma = 4$) & 100 & $-25.5\ \ (5.8)$ & $28.3\,(13.1)$ & $26.3\,(12.3)$ & $1.67\ (0.06)$ \\ \hline
Flat ($\tau = 0$) & 100 & $-21.0\ \ (6.6)$ & $16.7\ \ (4.3)$ & $15.1\ \ (4.0)$ & $1.66\ (0.09)$ \\ \hline
Mirror ($\tau = -1$) & 100 & $-35.7\,(12.5)$ & $54.0\,(11.1)$ & $50.3\,(10.5)$ & $2.06\ (0.14)$ \\ \hline
Scale ($\tau = 2$) & 100 & $-9.5\ \ (2.8)$ & $21.4\ \ (4.4)$ & $20.1\ \ (4.2)$ & $1.37\ (0.03)$ \\ \hline
Scale ($\tau = 1.29$) & 100 & $-1.5\ \ (0.8)$ & $3.1\ \ (0.8)$ & $2.9\ \ (0.7)$ & $1.13\ (0.03)$ \\ \hline
\end{tabular}
\end{table*}
\begin{table*}
\centering
\caption{Impact of undetected attacks despite deep aggregated monitoring. Results show median values over 365-day simulations together with their respective interquartile range. The participation rate is $\nicefrac{N}{M}=100\%$.}
\label{tab:attacks2}
\begin{tabular}{c|c|r|r|r|r}
\hline
\multirow{2}{*}{\textbf{Attack type}} & \multirow{2}{*}{\boldmath$\rho\ (\%)$} & \multicolumn{1}{c|}{\textbf{Attacker}} & \multicolumn{1}{c|}{\textbf{Other players}} & \multicolumn{2}{c}{\textbf{Utility company}}  \\
& & \textbf{Bill change (\%)} & \textbf{Bill change (\%)} & \textbf{Revenues change (\%)} & \textbf{PAR value} \\ \hline\hline
Shift ($\sigma = 4$) & 16 & $-0.8\ (0.7)$ & $1.1\ (0.5)$ & $1.0\ (0.5)$ & $1.22\ (0.11)$ \\ \hline
Flat ($\tau = 0$) & 28 & $-1.9\ (1.1)$ & $0.3\ (0.5)$ & $0.2\ (0.5)$ & $1.23\ (0.05)$ \\ \hline
Mirror ($\tau = -1$) & 16 & $-1.7\ (1.1)$ & $0.9\ (0.7)$ & $0.8\ (0.7)$ & $1.25\ (0.06)$ \\ \hline
Scale ($\tau = 2$) & 28 & $-1.0\ (0.7)$ & $2.3\ (0.7)$ & $2.2\ (0.7)$ & $1.11\ (0.04)$ \\ \hline
*Scale ($\tau = 1.29$) & 100 & $-1.5\ (0.8)$ & $3.1\ (0.8)$ & $2.9\ (0.7)$ & $1.13\ (0.03)$ \\ \hline
\multicolumn{6}{l}{* denotes attack that remains undetected even when applying deep individual monitoring}
\end{tabular}
\end{table*}

Table~\ref{tab:attacks1} reports the impact of undetected attacks despite average amount monitoring. As such monitoring is ineffective against the considered attacks, the attacker is able to carry out their attack with maximum strength, i.e. ($\rho = 100\%$), without being detected. The mirror ($\tau = -1$) attack is particularly efficient: The attacker's bill is reduced by $35.7\%$ at the cost of the other players' bills, i.e.~$54.0\%$, and a large increase of the PAR value to 2.06 from a non--attack value of 1.12.

Once deep aggregated monitoring is in place, the strength of the attacks that remain undetectable is reduced significantly. As Table~\ref{tab:attacks2} shows, the attacker's bill is lowered by 1.9\% at most. However, although, in this case, the other players are hardly affected - their bills only increase by 0.3\%, the UC suffers from a significant degradation of the PAR to 1.23. One should note that although the scale ($\tau=2$) attack with ($\rho=28\%$) produces a slightly better PAR value, i.e.~1.11 instead of 1.12 from the non--attack scenario, this is achieved by increasing the average electricity costs by 2.2\%.

Finally, although the most stringent monitoring strategy, i.e.~deep individual monitoring, would detect most attacks whatever $\rho$, i.e.~shift ($\sigma = 4$), flat ($\tau = 0$), mirror ($\tau = -1$) and scale ($\tau = 2$), some limited scale attacks such as ($\tau = 1.29,\ \rho = 100\%$) still cannot be discovered (cf.~last line of Table~\ref{tab:attacks2}). 
Although none of the proposed monitoring strategies can detect all attacks, they are able to recognise the most severe ones. Moreover, they can detect false data injection for a wide range of attacks.

\subsection{Attack Mitigation}
\label{attack_mitigation}
Once an attack has been detected, some response needs to be provided. For the most serious attacks, households may be instructed not to follow the calculated battery schedule, but use an alternative one. Several options are possible such as keeping the same schedule as the previous day or recalculating their schedule only taking into account their own data. In the latter case, scheduling is executed without using the game--theoretic framework, but by performing a simple optimisation of battery usage for their own consumption forecast.

Those options were evaluated in a previous study~\cite{Pilz2018a}. It showed that, although both approaches lead to a PAR reduction, local scheduling should be the defence of choice since it systematically outperforms previous day scheduling. Still, this mitigating strategy has its own cost: At medium participation rates $\nicefrac{N}{M}$, the PAR reduction can be up to~$\approx 25\%$ lower than when the game is played. 
As Tables~\ref{tab:attacks1} and~\ref{tab:attacks2} show, only the most powerful attacks have an impact on the PAR which is higher than reverting to the local scheduling strategy. This suggests that the best reaction to a low impact attack would be to let it happen. In terms of monitoring, only deep aggregated monitoring would prove useful, since it is able to detect all attacks for which the proposed mitigation strategy is beneficial. Therefore, a two-level detection system may be the most suitable strategy for the UC: It should conduct either no monitoring at all, or deep aggregated monitoring. 

Before deciding on a complete defence strategy, which includes detection and mitigation, all costs and benefits must be taken into account by the UC, i.e. cost of monitoring, cost of mitigating action, cost of reputation loss and benefit of increased consumption. The main challenge for the utility company is to control the spending on their security measures, as organisations typically have a restricted budget. For example, if the expected probability of an attack is low, a low investment in security could be justified. On the other hand, if an attacker is aware of such a strategy, they would be more likely to attack as they would expect less resistance. Finding a solution to this decision--making problem cannot be achieved by optimisation alone, but instead non-cooperative game theory helps in devising suitable models and advising on the expected likelihood of attacks.

\section{Game--Theoretic Defence Strategy}
\label{sec-game-defence}
When planning to defend against the false data injection attacks described in the previous section, the need for the utility company to allocate resources for the defence in the most efficient way has been highlighted. This section proposes to use game theory in order to support this decision--making process. 
The game is motivated and introduced based on detailing the payoff functions of the two players describing the game normal form. This is followed by solving the game using various assumptions. Finally, the solution is discussed with respect to their implications for the simulated scenario and potential alternatives.

\subsection{Game Theory for Security }
Game theory is increasingly being employed for modelling attacker-defender scenarios in cyber security, for a broad range of  scenarios such as intrusion detection in network security~\cite{Alpcan2010}, managing the security of information in an organisation~\cite{Panaousis2014CybersecurityApproach} and predicting the likelihood of cyber attacks~\cite{Bao2017HowGames}.

Non-cooperative game theory is based on the assumption that players are rational, i.e.~they choose between actions such that they maximise their payoffs. The associated \textit{optimal strategies} can be identified using the fundamental concept of the Nash Equilibrium (cf.~Section~\ref{sec:scheduling_game}). Although not all games have Nash Equilibrium, Nash's theorem states that nonzero-sum games always admit a mixed strategy equilibrium.
However, for practical applications it may not be easy to interpret~\cite{Arrow1985WhatAccomplish}.

In this paper, $x$ and $y$ denote a pure or mixed strategy of the first and second player in a two-player game, and $x^*$ and $y^*$ are used for optimal strategies of these players. A \textit{strategy profile} $s=(x, y)$ groups strategies of each player together. If the grouped strategies are optimal, the optimal strategy profile is written as $s^*$. A two-player nonzero-sum game can be represented in normal form, based on the players' payoff matrices $A$ and $B$~\cite{Shoham2009}.

An \textit{optimal Nash Equilibrium strategy profile} is a strategy profile $s^*=(x^*,y^*)$ satisfying  
\begin{equation}  
	x^*Ay^* \ge  xAy^* \;\;\forall x,\quad x^*By^* \ge    x^*By\;\;\forall y\ . 
\end{equation}
Here, the strategies may be pure or mixed, and the corresponding NE is referred to as pure or mixed. Furthermore, if all of the inequalities in the above definition are strict, one has a \textit{strict} NE. Otherwise, the NE is \textit{non-strict}.

\subsection{Proposed Security Game}
The proposed security game is a two-player nonzero-sum \textit{complete information} game~\cite{Shoham2009} between the utility company $\mathcal{U}$ and the attacker $\mathcal{A}$. 
The game is inspired by the nonzero-sum Intrusion Detection System (IDS) game of \cite{Alpcan2010} which has been thoroughly analysed in the literature and is well understood. Table~\ref{table:game_IDS} illustrates the game where the two strategies available to the defender are to monitor or not, denoted by the strategy space $S_\mathcal{D} = \{s^\mathcal{D}_{mon},s^\mathcal{D}_{-mon}\}$, and the attacker chooses between attacking or not attacking: $S_\mathcal{A} = \{s^\mathcal{A}_{att},s^{\mathcal{A}}_{- att}\}$. The positive parameters $\alpha_c, \alpha_f, \alpha_m, \beta_c$ and $\beta_s$ are used to denote the payoffs corresponding to the various strategies. The main characteristic of this game is the design of the payoff functions in such a way that the monitoring defender only has an incentive to defend in the presence of an attack. The attacker is discouraged from attacking if there is defence in place. This design leads to a circular path when considering payoff-incrementing unilateral changes of strategy, hence prohibiting the existence of a pure Nash Equilibrium.
\bgroup 
\def\arraystretch{1.5} 
\begin{table} 
	\centering 
	\caption{IDS Game of \cite{Alpcan2010} in Normal Form}\label{table:game_IDS} 
	\begin{tabular}{ c | cc | cc } 
    	$\mathcal{D} \downarrow\  \mathcal{A}\rightarrow$ & \multicolumn{2}{c|} {$s^\mathcal{A}_{att}$} & \multicolumn{2}{c}{$s^{\mathcal{A}}_{- att}$} \\\hline
        $s^{\mathcal{D}}_{mon}$ & $\alpha_{c}$, &$- \beta_{c}$ & $-\alpha_{f}$, &$0$\\\hline 
        $s^{\mathcal{D}}_{-mon}$ & $-\alpha_{m}$, &$\beta_{s}$ & $0$, &$0$ 
    \end{tabular}
\end{table}
\egroup 
\vspace{1em}
\subsubsection{Description of the Game}
Here, an augmented security game is introduced, extending the IDS game described previously by an additional action. The rationale behind this extended game model is twofold: Section \ref{sys_monitor} demonstrated the existence of low-impact attacks which cannot be detected by standard monitoring techniques, and it would be desirable to capture these in a more sophisticated game model. Second, an extended game might better match real-world scenarios and might lead to simpler solutions, in this case pure equilibria rather than mixed ones. 
\vspace{1em}
\paragraph{Game Strategies}
Section \ref{vulnerability_section} presented three possible monitoring strategies for $\mathcal{U}$: to monitor the daily average of forecasting data, to inspect the daily profile of the aggregated forecast, and to inspect the individual forecast data with the same level of detail. In this work, the assumption is made that the first and second monitoring strategies are most useful in a realistic setting, as they have an observable impact on the strength and outcome of successful attacks while the third monitoring strategy merely eliminates attacks that are possible for weaker monitoring levels. Furthermore, as the data of aggregated forecasts is readily available to the UC, the first monitoring strategy is not very costly and is identified with the strategy $s^\mathcal{U}_{-mon}$. The second monitoring strategy is denoted as $ s^\mathcal{U}_{mon}$ so that the strategy space for the defender $\mathcal{U}$ is as in the previous game $S_\mathcal{U} = \{s^\mathcal{U}_{mon},s^\mathcal{U}_{-mon}\}$. The attacker $\mathcal{A}$ has three different strategies: to attack strongly with high impact, to perform a weaker attack with low impact, or not to attack at all. This is denoted by the strategy space $S_\mathcal{A} = \{s^\mathcal{A}_{att{\degree}},s^\mathcal{A}_{att},s^{\mathcal{A}}_{- att}\}$. 

The additional weak attack strategy $s^\mathcal{A}_{att\degree}$ offers an alternative incentive of not monitoring to the UC, preferring to save monitoring cost when facing a weak attack. No assumption is made on the relationship between the attacker's overall payoff when choosing the two different attack types, and a discussion of conditions clarifying this relationship is the main subject of the game analysis in the next section.
\vspace{1em}
\paragraph{Game Payoff Functions}
The following notations for the payoffs for $\mathcal{U}$ are introduced: $c^\mathcal{U}_{mon}$ is the cost for monitoring  the daily profile of the aggregated forecast (second monitoring strategy) and $c^\mathcal{U}_{def}$ is an additional cost for investing in defence mechanisms such as actions discussed in Section \ref{attack_mitigation}. Losses from weak and strong attacks are denoted by $l^\mathcal{U}_{att\degree}$ and $l^\mathcal{U}_{att}$ respectively. The payoff functions corresponding to $\mathcal{A}$ are the benefits and costs associated with weak and strong attacks, denoted by $b^\mathcal{A}_{att\degree} $, $c^\mathcal{A}_{att\degree}$, and $b^\mathcal{A}_{att} $ and $c^\mathcal{A}_{att}$, respectively. 
\bgroup
\def\arraystretch{1.5}     
\begin{table*}         
	\centering         
    \caption{Security Game in Normal Form}\label{tab:game_NF}         
    \begin{tabular}{ c | cc | cc | cc }         
    	$\mathcal{U} \downarrow\  \mathcal{A}\rightarrow$ & \multicolumn{2}{c}{$s^\mathcal{A}_{att\degree}$} & \multicolumn{2}{|c|}{$s^\mathcal{A}_{att}$} & \multicolumn{2}{c}{$s^{\mathcal{A}}_{- att}$} \\\hline 
        $s^{\mathcal{U}}_{mon}$ & $-c^{\mathcal{U}}_{mon} - l^\mathcal{U}_{att\degree}$, &$l^\mathcal{U}_{att\degree} - c^\mathcal{A}_{att\degree}$ & $-c^{\mathcal{U}}_{mon} -c^\mathcal{U}_{def}$, &$- c^{\mathcal{A}}_{att}$ & $-c^\mathcal{U}_{mon}$, &$0$\\\hline     
        $s^{\mathcal{U}}_{-mon}$ & $-l^\mathcal{U}_{att\degree}$, &$l^\mathcal{U}_{att\degree} - c^\mathcal{A}_{att\degree}$ & $-l^{\mathcal{U}}_{att}$, &$l^{\mathcal{U}}_{att}- c^{\mathcal{A}}_{att}$ & $0$, &$0$     
    \end{tabular}     
\end{table*}     
\egroup 
The monitoring activity always leads to monitoring costs for $\mathcal{U}$. If there is no monitoring, $\mathcal{U}$ incurs losses $l_{att\degree}^{\mathcal{U}}$ and $l_{att}^{\mathcal{U}}$ due to weak and strong attacks. Otherwise, despite monitoring, weak attacks cannot be detected, hence there is a resulting loss $l_{att\degree}^{\mathcal{U}}$. Strong attacks however are detected and mitigated against through some countermeasures, preventing any losses but leading to a defence cost $c^\mathcal{U}_{def}$. Finally, if there is no attack, then the only arising nonzero payoff function involved is the monitoring cost for $\mathcal{U}$. The attacker $\mathcal{A}$ obtains a benefit $b^\mathcal{A}_{att\degree} $ from a weak attack, but has to invest in attack costs $c^\mathcal{A}_{att\degree} $. Similarly, the cost $c^\mathcal{A}_{att} $ arises from a strong attack, however the model assumes the lack of a benefit for $\mathcal{A}$ due to the UC's defence mechanism.
Using these notations, the proposed security game $\mathcal{G}$ can be represented in normal form as shown in Table~\ref{tab:game_NF}.
\vspace{1em}
\subsubsection{Game Assumptions}
In this section, assumptions on the relationship of the various cost and benefit functions, which are part of the game payoff matrices, are listed and justified. 
\vspace{1em}
\paragraph{Assumptions from the IDS Game}
\label{IDS_Assumptions}
The cost for missing an attack $\alpha_m={l}^{\mathcal{U}}_{att}>0$ is interpreted as losses from an attack that is not mitigated against, the false alarm cost $\alpha_f=c^{\mathcal{U}}_{mon}>0$ as an ongoing monitoring cost and the detection penalty $\beta_c=c^{\mathcal{A}}_{att}>0$ as the cost for the attacker to conduct a strong attack. The gain from detection $\alpha_c = -c^{\mathcal{U}}_{mon}-c^{\mathcal{U}}_{def} > 0$ is reformulated as necessary cost to monitor and to defend in order to prevent damage. In order to preserve the mixed equilibrium property of the security game given by $-\alpha_m < \alpha_c$ it is then assumed that this attack prevention cost is less than the actual incurring attack damage, i.e.~$c^\mathcal{U}_{mon} + c^\mathcal{U}_{def} < l^\mathcal{U}_{att}$. This assumption is natural: In a typical security game, the defender does not spend more on attack prevention than what they potentially loose from an attack. Finally, $\beta_s={{l}}^{\mathcal{U}}_{att}-c^{\mathcal{A}}_{att}>0$ is the difference between the benefit from an undetected attack and the attack effort. This expresses a similar principle as above, but this time applied to the attacker $\mathcal{A}$ who does not spend more on an attack than the expected gain from it. These assumptions can be referred to as the Security Game Assumptions.
\vspace{1em}
\paragraph{Augmented Security Game}
\label{Augmented_Assumptions}
The assumptions required for the  augmented security game are in parts inspired by those of the IDS game, and also motivated by the experimental results presented in Section \ref{vulnerability_section} which suggest that strong attacks require targeting more victims, i.e.~a bigger effort.

For a weak attack, the attacker receives a greater payoff than the cost of the attack, implying 
\begin{equation} 
	c^\mathcal{A}_{att\degree} <  l^\mathcal{U}_{att\degree}\ . 
	\label{a4} 
\end{equation}
It can also be assumed that the cost for launching a strong attack is higher than that for a weak attack since a higher number of households has to be attacked
\begin{equation} 
	c^\mathcal{A}_{att} > c^\mathcal{A}_{att\degree}\ . 
	\label{a5} 
\end{equation}
Finally, a strong attack leads to higher losses for the utility (cf.~Section~\ref{sec:outcomeForAttacker})
\begin{equation} 
	l^{\mathcal{U}}_{att}>l^{\mathcal{U}}_{att\degree}\ . 
    \label{a6} 
\end{equation}
Note that in order to aid the game analysis, an assumption made in this game is that the benefit of the attacker is equal to the loss of the defender, $b^\mathcal{A}_{att}=l_{att}^{\mathcal{U}}\ $ and $b^\mathcal{A}_{att\degree}=l_{att\degree}^{\mathcal{U}}\ $. 

\subsection{Game Analysis}
\label{sec:gameAnalysis}
In this section, analysis of the security game $\mathcal{G}$ reveals existence of several NE strategies. Following the study of practical examples, the relevance of these strategies are discussed so that they can be used to inform UC's security investments.
\vspace{1em}
\subsubsection{Optimal Nash Equilibrium Strategies}
To solve the augmented security game, three distinct cases are considered. This is based on discussing the second order difference 
\begin{equation}
\Delta=q_{att\degree}-q_{att}\ ,
\label{eqn:delta}
\end{equation}
where $q_{att\degree}=l^\mathcal{U}_{att\degree} - c^\mathcal{A}_{att\degree}$ and $q_{att}=l^\mathcal{U}_{att} - c^\mathcal{A}_{att}$ describe the net-benefit for the attacker in case of a weak and strong attack, respectively.
\vspace{1em}
\paragraph{Case 1 $(\Delta>0)$}
In this case, the existence of a unique pure NE for the game $\mathcal{G}$ can be asserted. The corresponding NE strategy is for the UC to not monitor, and for the attacker to carry out a weak attack. Due to the uniqueness property these solutions are globally optimal.

\begin{proposition}
If $l^\mathcal{U}_{att\degree} - c^\mathcal{A}_{att\degree} > l^\mathcal{U}_{att} - c^\mathcal{A}_{att}$, the game $\mathcal{G}$ admits a unique pure Nash Equilibrium strategy profile of the form $s^*=(s^\mathcal{U}_{-mon}, s^\mathcal{A}_{att\degree}) $ and the corresponding payoffs $s^*_{\mathcal{U}} = -l^\mathcal{U}_{att\degree} $ and $s^*_{\mathcal{A}} = l^\mathcal{U}_{att\degree} - c^\mathcal{A}_{att\degree} $ are globally optimal. \label{prop1}
\end{proposition}
\begin{IEEEproof}
First, it needs to be verified that when choosing the pure strategy profile $(s^\mathcal{U}_{-mon}, s^\mathcal{A}_{att\degree}) $, none of the two players benefits from a unilateral change of pure strategy.

Focusing on the UC, the change of strategy $s^\mathcal{U}_{-mon} \rightarrow  s^\mathcal{U}_{mon}$ diminishes its payoff since $-l^\mathcal{U}_{att\degree} > -c^\mathcal{U}_{mon}-l^\mathcal{U}_{att\degree}$ due to the assumption of a positive monitoring cost. Considering the attacker, the change $s^\mathcal{A}_{att\degree} \rightarrow  s^\mathcal{A}_{att}$ is not beneficial because of the main assumption $\Delta>0$ of this case. Finally, the change of strategy $s^\mathcal{A}_{att\degree} \rightarrow  s^\mathcal{A}_{-att}$ reduces the payoff due to Assumption~\eqref{a4}. Second, a careful inspection of the payoff functions of the remaining strategies of the game, together with the fact that the assumption of Case 1 implies $l^\mathcal{U}_{att\degree} - c^\mathcal{A}_{att\degree} >- c^\mathcal{A}_{att}$, shows that there is no other pure NE.
\end{IEEEproof}
\vspace{1em}
\paragraph{Case 2 $(\Delta<0)$}
Similarly to the IDS game, the augmented security game has the same property of circular paths when performing unilateral changes strategy with increasing payoffs, hence prohibiting the existence of any pure NE.
\bgroup     
\def\arraystretch{1.5}     
\begin{table*}         
	\centering      			
    \caption{Representative mixed strategy probabilities for Case 2 attacks based on simulations (cf.~Section~\ref{sec:attackImpactAnalysis})}
    \label{tab:game_mixed}         		
    \begin{tabular}{ r | c | c | c }
    	 & $p_{att\degree} = 36.3\%$ & $p_{att}=63.7\%$ & $p_{- att}=0\%$ \\\hline 
    	$p_{mon}=71.7\%$ & $26.0\%$ & $45.7\%$ & $0\%$\\\hline     
        $p_{-mon}=28.3\%$  & $10.3\%$ & $18.0\%$ & $0\%$
    \end{tabular}    
\end{table*}     
\egroup 
\begin{proposition}
If $l^\mathcal{U}_{att\degree} - c^\mathcal{A}_{att\degree} < l^\mathcal{U}_{att} - c^\mathcal{A}_{att}$, the game $\mathcal{G}$ admits no pure NE. \label{prop2} 
\end{proposition}
\begin{IEEEproof}
The proof of this proposition is done very similarly to that of Proposition 1 by comparing the changes in payoff, following a unilateral change of strategy.  It is clear that there is no pure NE in the game restricted to the attacker strategies $s^{\mathcal{A}}_{att}$ and $s^{\mathcal{A}}_{- att}$, as the resulting subgame is identical to the IDS game. When augmented by the weak attack strategy $s^\mathcal{A}_{att^\degree}$, two cases may arise, depending on which of the strategy changes  $s^\mathcal{A}_{att} \rightarrow s^\mathcal{A}_{att^\degree}$ or $s^\mathcal{A}_{att^\degree} \rightarrow s^\mathcal{A}_{att}$, starting from the initial strategy profile $(s^{\mathcal{U}}_{mon},s^\mathcal{A}_{att})$, lead to an increased payoff for the attacker. 

In the first case, one observes the additional sequence of strategy changes $s^{\mathcal{U}}_{mon} \rightarrow s^{\mathcal{U}}_{-mon}$, $s^\mathcal{A}_{att^\degree}\rightarrow s^\mathcal{A}_{att}$ and $s^{\mathcal{U}}_{-mon}\rightarrow s^{\mathcal{U}}_{mon}$ leading back to the original strategy profile. These changes entail increased payoffs due to the assumption of positive monitoring cost, the condition $\Delta<0$ and the Security Game Assumptions. In the second case, the unilateral payoff change joins the circular  path of the IDS game, from which the proof follows as shown earlier.
\end{IEEEproof}
\vspace{1em}
\paragraph{Case 3 $(\Delta=0)$}
In this last case, one derives the inequality $l^\mathcal{U}_{att\degree} - c^\mathcal{A}_{att\degree} = l^\mathcal{U}_{att} - c^\mathcal{A}_{att}>- c^\mathcal{A}_{att}$ as in Case 1 and obtains a similar but weaker result, as the pure NE is not strict. A formal proof of the following proposition is omitted as it can be done similarly as that of Proposition 1 since the same payoff deviations are involved.

\begin{proposition}
If $l^\mathcal{U}_{att\degree} - c^\mathcal{A}_{att\degree} = l^\mathcal{U}_{att} - c^\mathcal{A}_{att}$, the game $\mathcal{G}$ admits a unique pure non-strict Nash Equilibrium strategy profile of the form $s^*=(s^\mathcal{U}_{-mon}, s^\mathcal{A}_{att\degree}) $ and the corresponding optimal payoffs are $s^*_{\mathcal{U}} = -l^\mathcal{U}_{att\degree} $ and $s^*_{\mathcal{A}} = l^\mathcal{U}_{att\degree} - c^\mathcal{A}_{att\degree} $. \label{prop3}
\end{proposition}
\vspace{1em}
\subsubsection{Quantitative Examples}
Attacks discussed in Section \ref{vulnerability_section} are further analysed using the proposed augmented security game. In order to establish which case they correspond to, estimations of the sign of $\Delta$~\eqref{eqn:delta}  are performed using previous simulation calculations. More specifically, 
$b^\mathcal{A}_{att}$ and $b^\mathcal{A}_{att\degree}$ are represented by the values of the \`Attacker bill change' ($\gamma$ and $\gamma_{\degree}$), reported in Tables~\ref{tab:attacks1} and~\ref{tab:attacks2} respectively, multiplied by the actual amount of the bill $\lambda$, e.g.~$b^\mathcal{A}_{att}=l_{att}^{\mathcal{U}}=\gamma\cdot\lambda $. Moreover, assuming a linear relationship between the number of attacked players and the cost of an attack, $c^\mathcal{A}_{att}$ and $c^\mathcal{A}_{att\degree}$ can be expressed using the values of percentage of targeted players ($\rho$ and $\rho_{\degree}$) shown in Tables~\ref{tab:attacks1} and~\ref{tab:attacks2} respectively, e.g.~$c^\mathcal{A}_{att}=\rho\cdot\kappa$. As a consequence, an attack type corresponds to  Case 2, i.e. $(\Delta<0)$, iff the following inequality is satisfied:
\begin{equation}
\frac{\gamma_{\degree} - \gamma}{ \rho_{\degree}-\rho} > \frac{\kappa} {\lambda}\ . \label{a7}
\end{equation}
with Assumption~\eqref{a4} stating $\nicefrac{\gamma_{\degree}}{ \rho_{\degree}} > \nicefrac{\kappa}{ \lambda}$\ . 

Evaluations of attacks reported in Tables~\ref{tab:attacks1} and~\ref{tab:attacks2} show that Case 2 applies to the shift ($\sigma=4$), flat ($\tau=0$), mirror ($\tau=-1$) and scale ($\tau=2$) attacks.
Hence, for none of those attacks a pure NE exits and only mixed strategies can be offered. Using the mirror attack as an example, Equation~\eqref{a7} requires $0.41 > \nicefrac{\kappa}{\lambda}$ and Assumption~\eqref{a4} imposes $0.11 > \nicefrac{\kappa}{\lambda}$. 

Since the scale ($\tau = 1.29$) attack was especially designed to be undetectable by the proposed monitoring solution, it cannot be analysed by the game which assumes that a successful monitoring strategy is available. On the other hand, the best strategy for such attack is self-evident: Since all attack results in gains for the attacker, they should attack, while the UC should not waste any resources in ineffective defence. 

In order to investigate the mixed strategies associated to those attacks, numeral values were selected so that mixed strategies could be computed using an NE solver~\cite{Avis2010EnumerationGames}: $\lambda = 100$, $\kappa = 10$, $c^{\mathcal{U}}_{mon} = 10$ and $c^{\mathcal{U}}_{def} = 20$. 
Table~\ref{tab:game_mixed} shows representative  mixed strategy probabilities associated with the investigated Case 2 attacks, here the mirror attack. The attacker either performs a strong (63.7\% probability) or weak (36.3\% probability) attack, while the UC chooses to use monitoring with a 71.7\% probability. Note that the choice of numerical values is not critical. As long as all the game assumptions are fulfilled, the probability for the monitoring action of the UC is at least 70\%.
\vspace{1em}
\subsubsection{Discussion}
Theoretical analysis of the proposed extended game model has shown that according to the sign of $\Delta$~\eqref{eqn:delta}, three different cases should be considered. While, both Case 1 $(\Delta>0)$ and Case 3 $(\Delta=0)$  are associated to a pure NE, only Case 1's is strict. However, in both cases, the optimal NE strategy for the UC is the same: not to monitor. On the other hand, Case 2 $(\Delta<0)$ only leads to mixed strategies. Practical analysis, investigating the attack examples described in Section \ref{vulnerability_section} based on a 100\% participation rate, revealed that only Case 2 was practically relevant. This is in line with expections that the net benefits, i.e. benefits minus costs, of strong attacks are supposed to be higher than those of weak attacks. Note that for the scale ($\tau=2$) attack, different cases could arise at lower participation rates due to its specific behaviour as shown in Fig.~\ref{fig:benefit_attacker} and Fig.~\ref{fig:attackExplanation}.

Regarding Case 2, for a UC, the practical application of optimal strategies, as illustrated in Table~\ref{tab:game_mixed}, is not straightforward. Actually many suggestions have been made regarding possible interpretations of mixed strategies~\cite{Aumann1985WhatAccomplish,Aumann1995EpistemicEquilibrium,Chen2017PowerApproaches}. In the specific context of this work, that proposed by~\cite{Chen2017PowerApproaches} is of particular interest: Indeed, assuming that the UC supplies a set of microgrids, where security strategy is decided at the microgrid level, they, seen as a population, would choose defence strategies following the mixed probabilities. Alternatively, as suggested in \cite{MaghrabiImprovedTheory,Shoham2009}, the probability associated to defence could be interpreted as an index of security criticality which would inform the UC's decisions regarding its defence investments. Interestingly, experiments (not shown) indicates that when the cost of attacking a singler player, i.e.~$\kappa$, decreases, the mixed strategy probability for monitoring grows, increasing defence needs. 

Finally, the undetectable scale ($\tau = 1.29$) attack is a reminder that no practical monitoring strategy is perfect and the best defence strategy may be not to defend if the losses associated to an attack can be considered as acceptable.


\section{Conclusion}
\label{sec:conclusion}
Protecting smart grids from cyber attacks is essential for them to deliver their promises. 
Investigating different classes of false data injection attacks against the forecasts required for smart energy scheduling, extensive simulations showed the extent of damages that a single attacker can cause to both the utility company (growth of PAR value by up to 84\%) and its consumers (bill increase by up to 54\%). The need for mitigation having been established, monitoring and defence strategies were proposed. In order to assess their value and advise utility companies on their optimal attack prevention strategy, a novel and generic security game that considers low and high-impact attacks was designed. Its analysis highlighted, in particular, conditions under which a Nash Equilibrium exists. Interestingly, in those cases, the best strategy is for the utility company not to invest in any monitoring and the attacker to conduct low-impact attacks. Numerical evaluations considering the previously studied classes of attacks revealed that there is a type of attack where, indeed, no monitoring is the best strategy. However, in all the other cases, only mixed strategies can be offered. Their practical interpretation by UCs was discussed. As a conclusion, the proposed security game offers utility companies the ability to investigate the most appropriate monitoring and defence strategies so that false data injection attacks have only very limited, if any, impact on smart energy scheduling.  

\section*{Acknowledgment}
This work was supported by the Doctoral Training Alliance (DTA) Energy, the Erasmus+ Programme of the European Union and the Kingston University First Grants Scheme.

\addtolength{\textheight}{-6cm}

\bibliographystyle{IEEEtran}

\appendices
\section{Simulating Forecasting Errors}
\label{sec:forecasting}
Since forecasting electricity consumption is out of the scope of this study, forecasts were simulated instead of produced by a forecasting algorithm. However, in order to consider forecasts as realistic, they must show some deviation from the actual consumption. As it has been reported that the average error in individual forecasted data is around 8\%~\cite{Bichpuriya2016}, some random error is added to the actual consumption values to produce sufficiently inaccurate forecasts. Although errors could be added following a Gaussian law, the obtained forecasted profile would prove unrealistic since it would display random jumps. As a consequence, some smoothing effect is added by linking successive values. More specifically, for each value $i$, a random error is initially calculated $e_i$, then the actual error added to the value $i$ is the average of the corresponding $e_i$ and its neighbours, i.e. $E_i = \frac{e_{i-1} + e_i + e_{i+1}}{3}$ .
\begin{figure}
	\centering
	\includegraphics[width=0.42\textwidth]{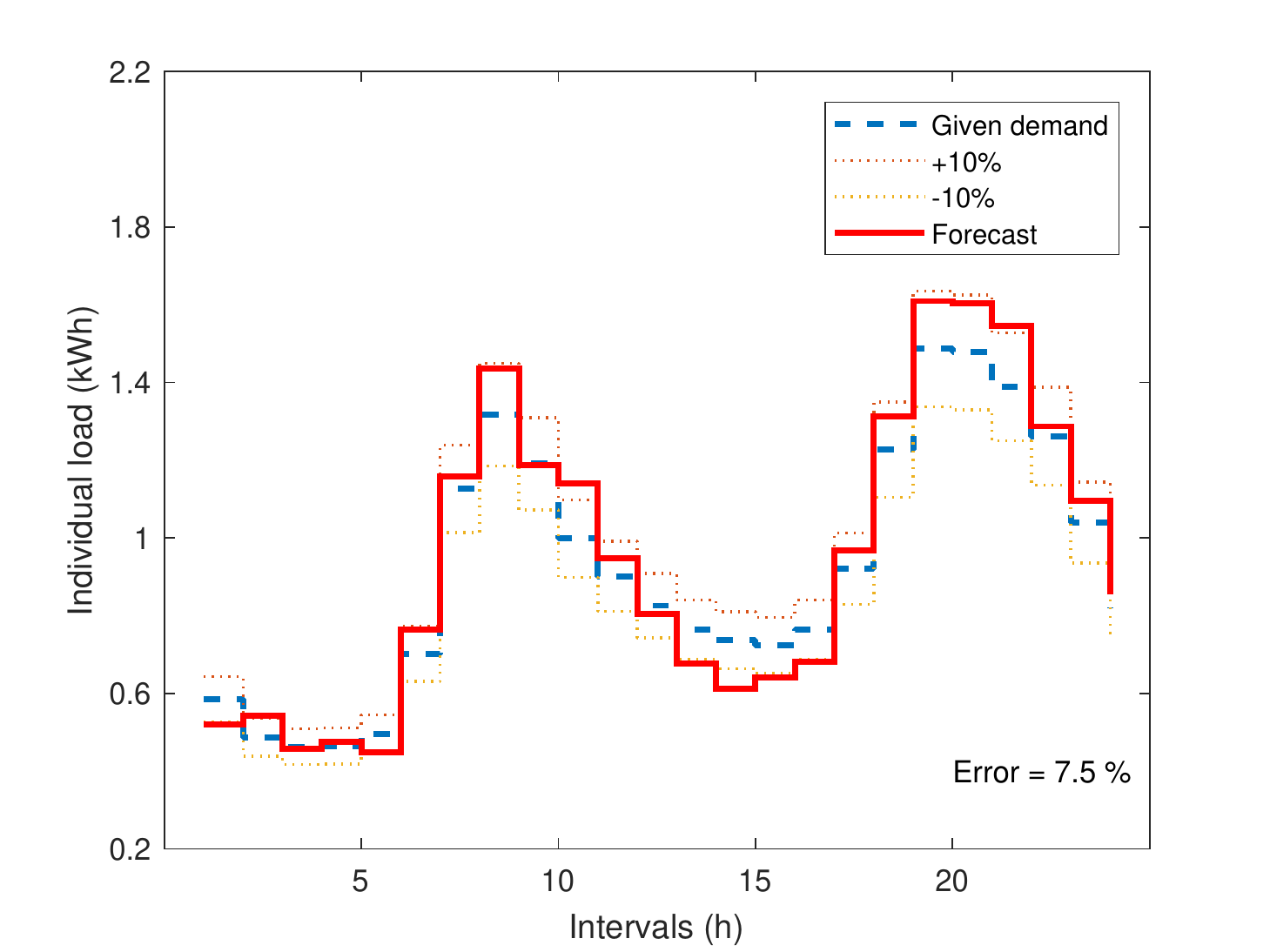}
	\caption{\textit{Individual forecast created by adding random errors.} While the dashed curve is the actual demand of an household, addition and subtraction of $10\%$ are represented by the two dotted curves. The bold curve is one example of simulated forecast produced using the described method. Here, whereas the average error is $7.5\%$, there are some values outside the $10\%$ error area.}
	\label{fig:individual_forecasting_errors}
\end{figure}
	As seen on Fig. \ref{fig:individual_forecasting_errors}, with this approach, simulated forecast is smoother and, as a consequence, more realistic. Due to the relatively large number of players, despite the added errors, the aggregated forecast remains quite similar to the aggregated demand (an average error of around 2\% was estimated experimentally). As a consequence, game solutions based on forecast with and without errors are close: drawing the histogram of the error per day during a whole year (not shown) reveals an average error of 8\%~\cite{Pilz2018AErrors}. 
\addtolength{\textheight}{8cm}
\section{Supplementary material}
\label{sec:supplementary}
\begin{figure*}
	\centering
	\includegraphics[width = 0.75\textwidth]{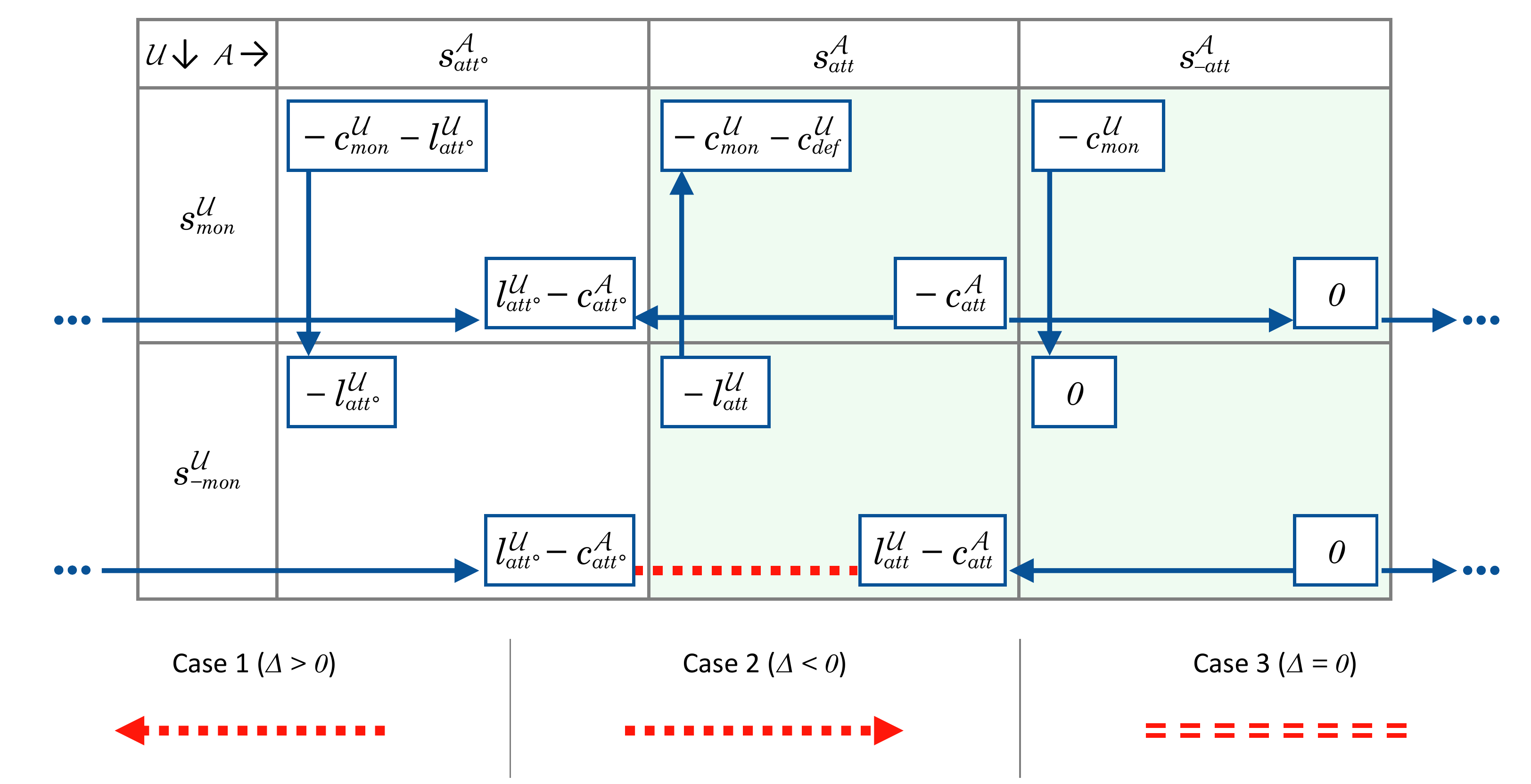}
	\caption{\textit{Advanced security game flow diagram.} This figure is a more extensive representation of the game shown in Table~\ref{tab:game_NF}, including the relations between the respective quantities. The arrows indicate which strategy would be more preferable in terms of the individual players' utility function. As discussed in Section~\ref{sec:gameAnalysis} the connection between the IDS game (in green) and the proposed augmented security game is defined by the second order difference $\Delta$~\eqref{eqn:delta} which is highlighted here by the red dotted lines. Depending on the sign of $\Delta$~\eqref{eqn:delta}, the direction of the arrows varies as illustrated in the three cases. Note that the double line represents equality.}
	\label{fig:game_analysis}
\end{figure*}

Fig.~\ref{fig:game_analysis} shows a flow diagram of the augmented security game which helps to understand the analysis in Sec.~\ref{sec-game-defence}.

\begin{figure*}
	\centering
	\includegraphics[width = 0.80\textwidth]{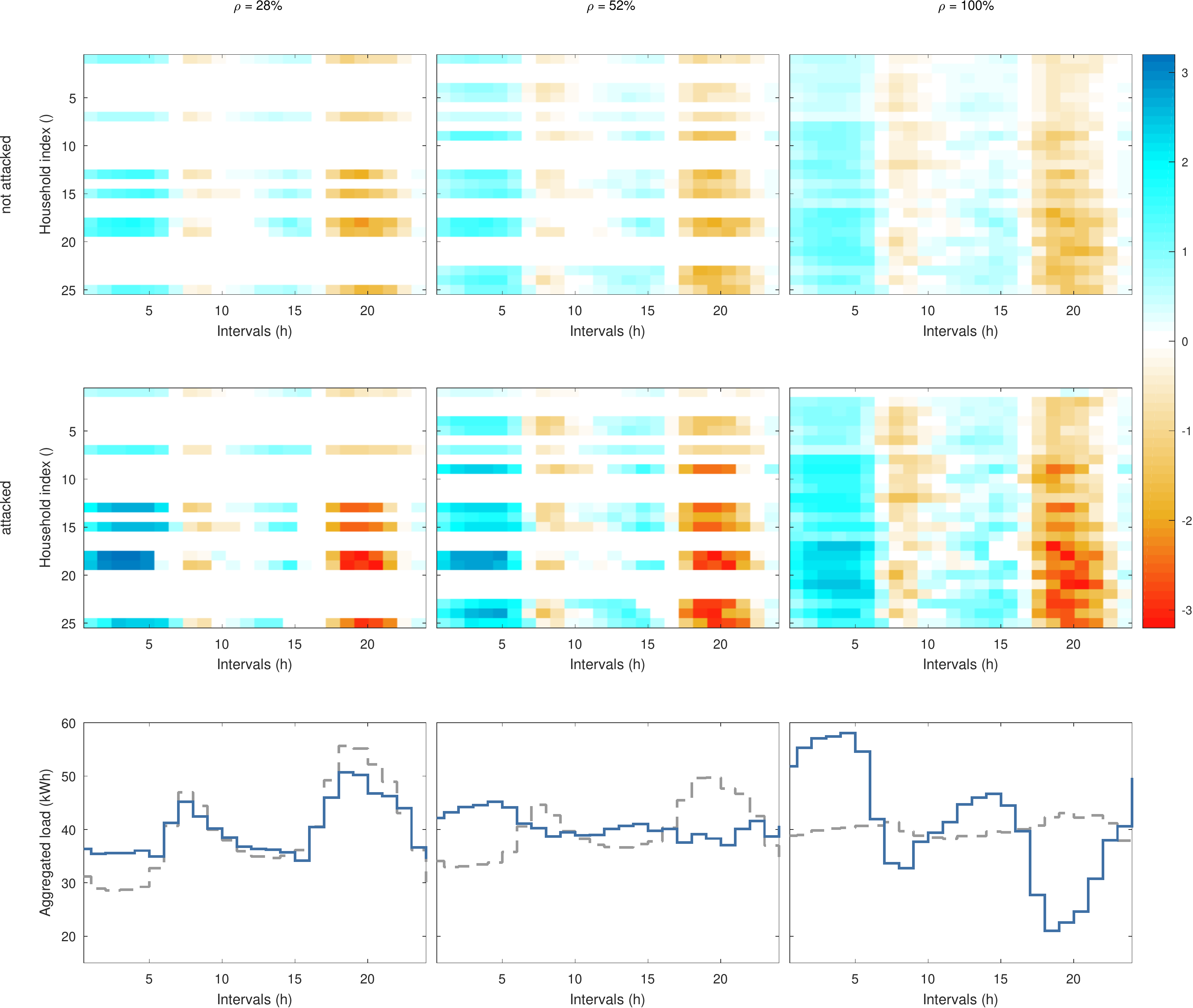}
	\caption{\textit{Aggregated load and battery schedule without and under a scale ($\tau=2$) attack targetting all players for different household participation rates ($\rho$)}. Each column corresponds to a different participation rate, i.e. from left to right $\rho=28\%$, $\rho=52\%$ and $\rho=100\%$. The first row shows battery schedules of each individual household; the second row shows battery schedules of each individual household under attack - note that the first household is the attacker; the third row compares aggregated loads without - dashed curves - and with - bold curves - attacks. Without attack, participation of all households, i.e.~$\rho=100\%$, is required to flatten the aggregated load ($\text{PAR}=1.07$). However, excessive battery usage by attacked households (the second row shows stronger charges and discharges) leads to a relatively flat ($\text{PAR}=1.11$) aggregated load at $\rho=52\%$. However, at $\rho=100\%$ the aggregated load profile is almost inverted; in this case the attacker hardly needs to use their battery.}
	\label{fig:attackExplanation} 
\end{figure*}

Fig.~\ref{fig:attackExplanation} provides details to the discussion in Sec.~\ref{sec:outcomeForAttacker} about individual household schedules and the influence of the scale attack with $\tau=2$. 

\ifCLASSOPTIONcaptionsoff
  \newpage
\fi

\end{document}